\documentclass[reprint,graphicx,amsmath,amssymb]{revtex4-1}
\usepackage{url}
\usepackage{graphicx}
\def\eqnn#1{Eq.~(\ref{eq:#1})}

\def\figno#1{Fig.~\ref{fig:#1}}
\def\secno#1{Sect.~\ref{sec:#1}}

\def\vev#1{\langle#1\rangle}

\def\method#1{{\it#1}}
\begin{document}
\title{
  Noise reduction and hyperfine level coherence 
  in spontaneous noise spectroscopy of atomic vapor
}
\author{Takahisa Mitsui and Kenichiro Aoki}
\affiliation{
Research and Education Center for Natural Sciences and
   Hiyoshi Dept. of Physics,
   Keio University, Yokohama 223--8521, Japan}
\begin{abstract}
    We develop a system for measurements of power spectra of
    transmitted light intensity fluctuations, in which the extraneous
    noise, including shot noise, is reduced.  In essence, we just
    apply light, measure the power  of the transmitted light and
    derive its power spectrum.
    We use this to observe the spontaneous noise spectra of photon
    atom interactions.
    Applying light with frequency modulation, we can also observe the
    spontaneous noise reflecting the coherence between the hyperfine
    levels in the excited state.
    There are two main novel components in the measurement system, the
    noise reduction scheme and the stabilization of the laser
    system. The noise reduction mechanism can be used to reduce the
    shot noise contribution to arbitrarily low levels through
    averaging, in principle.  This is combined with differential
    detection to keep unwanted noise at low levels.  The laser system
    is stabilized to obtain spectral width below 1\,kHz without high
    frequency ($\gtrsim10\,$MHz) noise.  These methods are described
    systematically and
    the performance of the measurement system is examined through
    experimental results.
\end{abstract}
\maketitle
\section{Introduction}
To study the nature of photon atom interactions, a simple experiment
would be to just shine light on atoms and measure the fluctuations in
the transmitted light. We call these fluctuations ``spontaneous
noise'' since they arise solely from photon atom interactions, {\it
  without } any other external
perturbations\cite{AZ,Mitsui2,Crooker2004}. By measuring its spectra
over a wide frequency range and analyzing its structure, we may obtain
a comprehensive picture of photon atom interactions.  Transmitted
light has a large intensity when compared to fluorescence, making it
possible to measure details of the spontaneous noise spectra.

There is a well known obstacle that needs to be overcome in achieving
such measurements; namely, the shot noise that inevitably arises in
the photon flux. The shot noise appears as white noise in the current
fluctuation spectrum, through photoconversion, as
\begin{equation}
    \label{eq:shotNoise}
    \vev{\left(\Delta I\right)^2}    = 2eI\Delta f    \quad.
\end{equation}
Here, $e$ is the elementary charge, $I$ the electric current and $f$,
the frequency. As explained below, we reduce this noise statistically,
while retaining the spontaneous noise, to obtain its spectra below
shot noise levels. In practice, there will be other unwanted noise
that occur and we shall also need to eliminate them.  The two main
components of the measurement system are the noise reduction scheme
and the laser stabilization.  This measurement system, which combines
the averagings of correlations and differential detection was first
used in \cite{amRb}, leading to  high contrast observations of
spontaneous noise spectra.  A main objective in this work is to
describe the principles underlying this system and to investigate its
efficacy through the systematic analysis of experimental results. The
measurement system is relatively simple and easy to set up, so that we
believe that there is some merit in explaining this system for future
use. Another main objective here is to report on the observations of
spontaneous noise arising from hyperfine levels with coherence, which
has not been seen previously.

To reduce shot noise in measurements, several approaches are
conceivable: If one is trying to measure a signal with a known
frequency, one can make measurements at this frequency to reduce the
shot noise, by effectively making $\Delta f\longrightarrow 0$ in
\eqnn{shotNoise}. However, such method precludes obtaining spectra
over a wide frequency range, which is our objective. Another approach
is to use a squeezed light source which has smaller shot noise from
the start\cite{quantumOptics,Caves81,squeezed}. While this approach is elegant and has yielded shot
noise reduction by a factor of two\cite{squeezeSource}, it seems so far
to be difficult to use this to reduce the shot noise by orders of
magnitude.
Yet another approach to reduce shot noise, relatively, is to increase
the intensity of the light source, since the intensity power spectrum
behaves as the square of the intensity and the shot noise behaves as a
single power, as seen in \eqnn{shotNoise}.  Such approach is not
applicable to the spontaneous noise spectra measurements, since
changing the light intensity changes the properties of the spontaneous
noise itself, which is what we wish to study.

In this work, we use a coherent light source and reduce shot noise
through averagings by a factor of $10^{-3}$.  This noise reduction
principle is simple and has a wide range of applicability; it has
proven effective in obtaining surface thermal fluctuation spectra of
liquids, complex fluids and biological material\cite{am1,am2}.

In \secno{setup}, the noise reduction scheme we employ is
systematically explained and the methods used to stabilize the laser
system are described in \secno{laser}. The performance of these methods
are examined in the measurements of spontaneous noise spectra in
\secno{exp}. In \secno{hfCoherence}, spontaneous noise arising from
Rabi flopping between hyperfine levels coupled through the frequency
modulation of the applied light is investigated. \secno{conc} contains
conclusions and discussion.
\section{Noise reduction scheme and the experimental setup}
\label{sec:setup}
\begin{figure}
    \centering
    \includegraphics[width=8.4cm,clip=true]{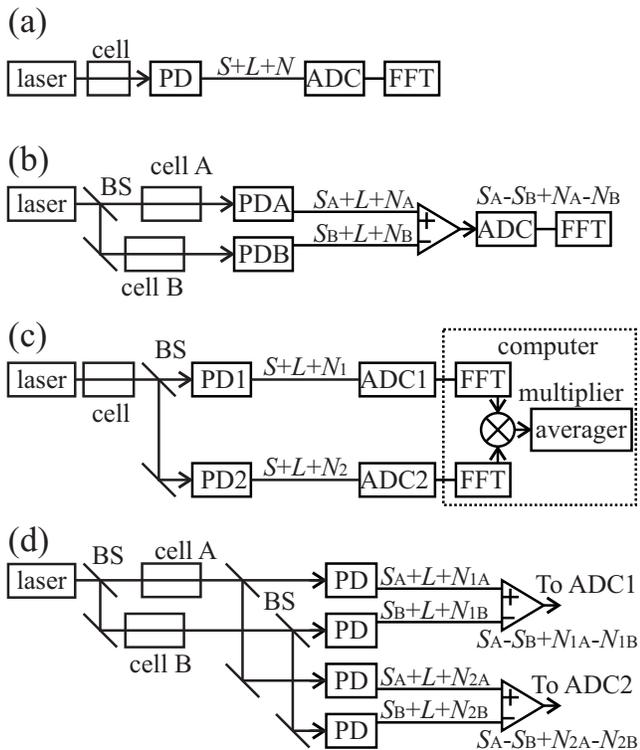}
    \caption{{The schematics for the measurement systems: } (a) A
      straightforward scheme for measuring intensity fluctuations of
      light passing through a vapor cell. (b) Measurement system with
      differential detection. (c) A measurement system with shot noise
      reduction. (d) A measurement system with both differential
      detection and shot noise reduction. BS: Beam splitter.
      FFT: Fast  Fourier transform. }
    \label{fig:setup}
\end{figure}

Let us briefly outline the principles underlying the noise reduction
method used in the experiment, before explaining this in more detail.
In the experiment, we measure the power spectrum of the transmitted
light intensity.
To reduce various extraneous noise that inevitably occur in this
measurement while keeping the spontaneous noise intact, we partition
the photons in each transmitted light beam to allow for correlation
measurements.  These measurements contain extraneous noise, both
correlated and uncorrelated, in addition to the spontaneous noise,
which we want to measure. We shall remove this extraneous noise by
combining statistical methods and differential detection. The
experimental setup with varying degrees of noise reduction are shown
in \figno{setup}(a)--(d) and are correspondingly explained in the
following subsections {\it a--d}. They will subsequently  be referred
to  as systems \method{a}--\method d.
\paragraph{The basic concept of the measurement:}
Conceptually, the measurement of the power spectrum of spontaneous
noise is straightforward and is indicated in \figno{setup}(a).  A
laser light is shone on a cell containing atomic vapor. The
transmitted light intensity is converted to an electric current by a
photodetector (PD), which is then converted to a digital signal
through an analog-to-digital converter (ADC) and Fourier transformed
to obtain the spectrum.

The shortcoming
of this setup is that unwanted noise can not be reduced enough to
obtain the spectrum of the spontaneous noise, $S$. Two main kinds of
such noise exist. First is the shot
noise\cite{quantumOptics}, 
$N$, which is a quantum fluctuation of the photon number that leads to
the noise in the photocurrent as
\eqnn{shotNoise}\cite{Mitsui1}. 
Second is the noise $L$ induced by the laser noise.  Some amplitude
modulation (AM) and frequency modulation (FM) noise exist in the laser
system, even when care is taken to minimize them, as described below.
In particular, the effects of FM noise are not negligible even when
the spectral width is narrowed using an external cavity.

Denoting the measured photocurrent at time $t$ as $D_{\rm
  a}(t)=S(t)+L(t)+N(t)$, its power spectrum is
\begin{equation}
    \label{eq:spectrumA}
    \vev{|\tilde D_{\rm a}(\omega)|^2}=
    \vev{|\tilde S(\omega)|^2}+
    \vev{|\tilde L(\omega)|^2}+
    \vev{|\tilde N(\omega)|^2}
    \qquad.
\end{equation}
Here, $\vev{\cdot}$ denotes averages, tildes denote Fourier transforms
and $\omega$ is the (angular) frequency. We used the property that
$S(t),L(t),N(t)$ arise from different physics and hence are
uncorrelated in this setup. In the measurements, the magnitudes of
these three noises are comparable and there are no distinct criteria
for separating them.  So
it is not possible to extract the spontaneous noise spectrum 
$\vev{|\tilde S(\omega)|^2}$ with this method. 
\paragraph{Differential detection:}
In \figno{setup}(b), a measurement system using two independent vapor
cells is shown. Since $S$ is generated by atoms spontaneously, signals
arising in the two separate atomic systems (cells A, B) are
statistically uncorrelated, even if they are set up identically. On
the other hand, $L$, caused by a single laser light source is identical
in these two measurements. This allows us to eliminate laser noise
effects using differential detection. The measured photocurrent using
differential detection is $D_{\rm b}(t)=S_{\rm A}(t)-S_{\rm
  B}(t)+N_{\rm A}(t)-N_{\rm B}(t)$, where the suffixes A, B refer to
the two separate vapor cells. Since $S_{\rm A,B},N_{\rm A,B}$ are all
uncorrelated, the power spectrum is
\begin{eqnarray}
    {1\over2}\vev{|\tilde D_{\rm b}(\omega)|^2}&=&
    {1\over2}\Bigl(
      \vev{|\tilde S_{\rm A}(\omega)|^2}+    \vev{|\tilde S_{\rm
          B}(\omega)|^2}
      \\&&\nonumber\qquad
      +
      \vev{|\tilde N_{\rm A}(\omega)|^2}+    \vev{|\tilde N_{\rm B}(\omega)|^2}
    \Bigr)\nonumber\\
    &=    &
    \label{eq:spectrumB}
    \vev{|\tilde S(\omega)|^2}+    \vev{|\tilde N(\omega)|^2}
    \qquad.
\end{eqnarray}
Since cells A, B contain atomic vapor with identical properties, $
\vev{|\tilde S_{\rm A}(\omega)|^2}= \vev{|\tilde S_{\rm
    B}(\omega)|^2}= \vev{|\tilde S(\omega)|^2}$. Shot noise spectra
for A, B are also equal.  While it is clear that the laser noise
induced fluctuations $L$ can be eliminated with this method, effects
due to shot noise $N$ still remain, since they are uncorrelated in 
PD A, B.
Therefore, another method is required to extract the spontaneous noise
spectrum.
\paragraph{Shot noise reduction:}
While shot noise is inherent in any light source, it is random and its
properties are universal. Therefore, we can use correlation
measurements to remove it\cite{am3}. 
In the setup  of \figno{setup}(c), the laser light passing through the
vapor cell is partitioned into two and measured by the two photodetectors,
PD1, PD2. By Fourier transforming the current signals from PD1, PD2,
$D_{{\rm c}j}(t)=S(t)+L(t)+N_j(t)\ (j=1,2)$, and taking their
correlation, we obtain
\begin{equation}
    \label{eq:spectrumC}
    \vev{\overline{\tilde D_{{\rm c}1}(\omega)}\tilde D_{{\rm c}2}(\omega)}
    \longrightarrow \vev{|\tilde S(\omega)|^2}+    \vev{|\tilde
      L(\omega)|^2}\quad,
\end{equation}
in the limit of infinite number of averagings.  In the formula, bar
denotes complex conjugation. Here, we used the
property that $\tilde S(\omega)$, $\tilde{L}(\omega)$,
$\tilde{N}_1(\omega)$, $\tilde{N}_2(\omega)$ are uncorrelated in the
measurements.  Denoting the number of averagings, ${\cal N}$, the
experimental uncertainty in \eqnn{spectrumC} is statistical and its
relative size is $1/\sqrt{\cal N}$. The crucial element here is the quantum
property that the shot noise $N_{1,2}$ measured in the two independent
photodetectors are statistically uncorrelated, even if they occur
in the partitions of a single laser light source. 
It should be noted that this property holds for a coherent light
source, but not necessarily for a source with different photon
statistics\cite{quantumOptics}. 
This correlation measurement allows us to achieve shot noise reduction
to arbitrary levels, yet $ \vev{|\tilde L(\omega)|^2}$ still remains
since the laser noise originates from a single light source and are correlated.
Since it is impossible to measure just $\vev{|\tilde L(\omega)|^2}$,
which is induced in the atoms by the laser noise, further improvements
are necessary to obtain $\vev{|\tilde S(\omega)|^2}$.
It should be noted that this correlation measurement removes any
noise uncorrelated in $D_{c1},D_{c2}$, such as instrumental noise,  in
addition to shot noise.
\paragraph{Differential detection and shot noise reduction:}
To remove both $L$ and $N$, we combine the above two noise reduction
schemes effectively.  Shown in \figno{setup}(d) is the experimental
setup used in this work, which incorporates both the differential
detection and the shot noise reduction. The differentially detected
photocurrents sent to ADC1,2 are $D_{{\rm d}j}(t)=S_{\rm A}(t)-S_{\rm
  B}(t)+N_{j\rm A}-N_{j\rm B}\ (j=1,2)$
and the
correlation of their Fourier transforms is
\begin{eqnarray}
    \label{eq:spectrumD}
    {1\over2}\vev{\overline{\tilde D_{{\rm d}1}(\omega)}
      \tilde D_{{\rm d}2}(\omega)}
    &=&{1\over2}\left(
      \vev{|\tilde S_{\rm A}(\omega)|^2}+    \vev{|\tilde S_{\rm B}(\omega)|^2}    
      \right)\nonumber\\&=&
      \vev{|\tilde S(\omega)|^2}
    \qquad.
\end{eqnarray}
It can be seen that the spontaneous noise spectrum can be extracted,
without the shot noise nor the fluctuations induced by the  laser noise.
\section{Stabilization of  the laser system}
\label{sec:laser}
  \begin{figure}[htbp]
      \centering
    \includegraphics[width=8 cm,clip=true]{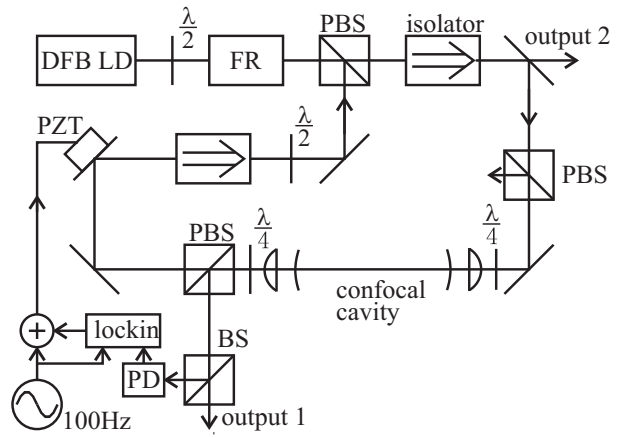}    
    \caption{The schematics of the stabilized laser system used in the
      experiment. $\lambda/2, \lambda/4$ indicate half- and
      quarter-wave plates.
      lockin: Lock-in amplifier. FR: Faraday rotator. PBS: Polarizing
      beam splitter. PZT: Piezoelectric transducer.  }
\label{fig:laser}
\end{figure}
A necessary part of the measurement system is a highly stabilized
light source with low levels of both AM and FM noise. We explain in this
section what is required and how to accomplish these requirements.
In this experiment, we used a semiconductor laser, DFB-0780-080,
Sacher lasertechnik, on a vapor of rubidium (Rb) atoms. The laser has
a wavelength of 780\,nm and a spectral width of about 3\,MHz, which is
of the same order as the spectral width of Rb-D$_2$ transitions, about
6\,MHz.  This FM noise induces fluctuations in the photon absorption
by Rb atoms\cite{Mitsui1},
whose magnitude is much larger than that of the spontaneous noise we
aim to extract.
The most effective method for stabilizing a semiconductor laser and
obtaining a narrow spectral width is to optically couple a confocal
cavity to it. We have been able to stabilize a distributed feedback
(DFB) laser diode (LD) to narrow down the spectral width below 1\,kHz,
using a previously proposed method\cite{Dahmani}. However, with this
approach, laser instability induced noise arises at the same levels as
the shot noise, so that it was impossible to extract the spontaneous
noise spectrum over a wide frequency range. One reason for this is
that in this method, the confocal cavity is operated off axis, so that
spurious modes tend to arise in the emitted light. Another reason is
that the method is ineffective in suppressing FM noise at higher
frequencies ($f\gtrsim10\,$MHz). Since semiconductor lasers contain FM
noise up to few GHz frequencies\cite{Yamamoto}, another method is
needed to achieve a stabilized laser system.

The stabilized laser system used in this experiment is shown in
\figno{laser}.  While the optical system is somewhat complex, in
essence, the light transmitted {\it on axis} through the confocal
cavity (SA--300, Technical Optics) is fed back into the laser diode,
which does not give rise to spurious modes and facilitates
adjustments. However, similarly to the previously proposed
method\cite{Dahmani}, the suppression of FM noise at high frequencies
is insufficient in the traditional output from the system (output 2 in
\figno{laser}). Therefore, we used light transmitted through the  
confocal cavity (output 1) to obtain the spontaneous
noise spectrum in this experiment.
The FM noise, however, can be used to generate interesting physics
effects, which we shall explore in \secno{hfCoherence}.
With our experimental setup, laser noise induced effects at
frequencies above 10\,MHz were undetectable when using output 1.
The use of light passed through the confocal cavity as in output 1 and
the coupling of the confocal cavity to the light source {\it on axis}
are new, we believe, and can be instrumental in achieving precision
measurements at low noise levels.
Light power of DFB LD was 60\,mW, while that of output 1
(\figno{laser}) was 2\,mW.  So, one aspect of this method is that we
have traded in some of the power for achieving a source with lower
noise levels.
Since output 1 is an output from a confocal
cavity, FM noise for frequencies below 10\,MHz is converted to AM
noise. Therefore, AM noise is increased at lower frequencies in output
1. So, output 2 (\figno{laser}) can be useful for spectral
measurements at lower frequencies and it also has a higher power than
output 1.
\section{Experimental results and the performance 
  of the measurement system} 
\label{sec:exp}
\def\hnr{8.5}
\begin{figure}
    \centering
    \includegraphics[width=\hnr cm,clip=true]{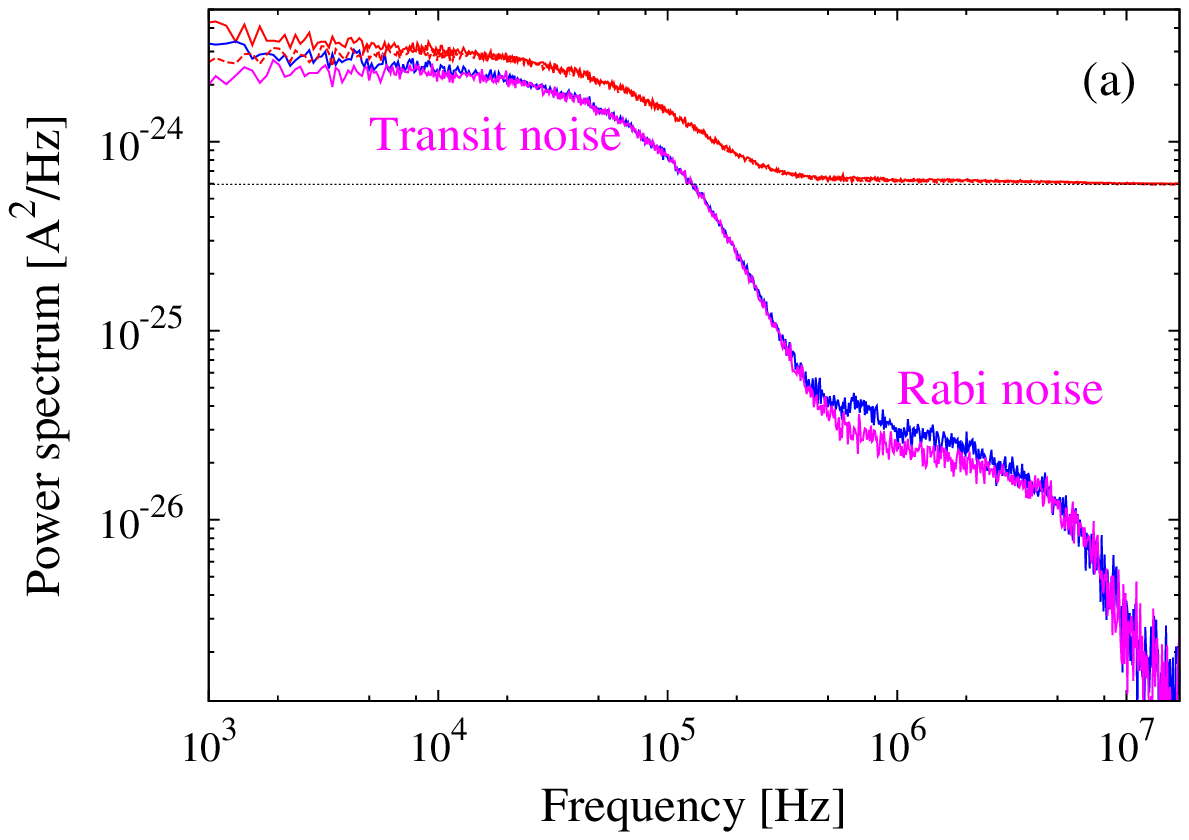}\\
    \includegraphics[width=\hnr cm,clip=true]{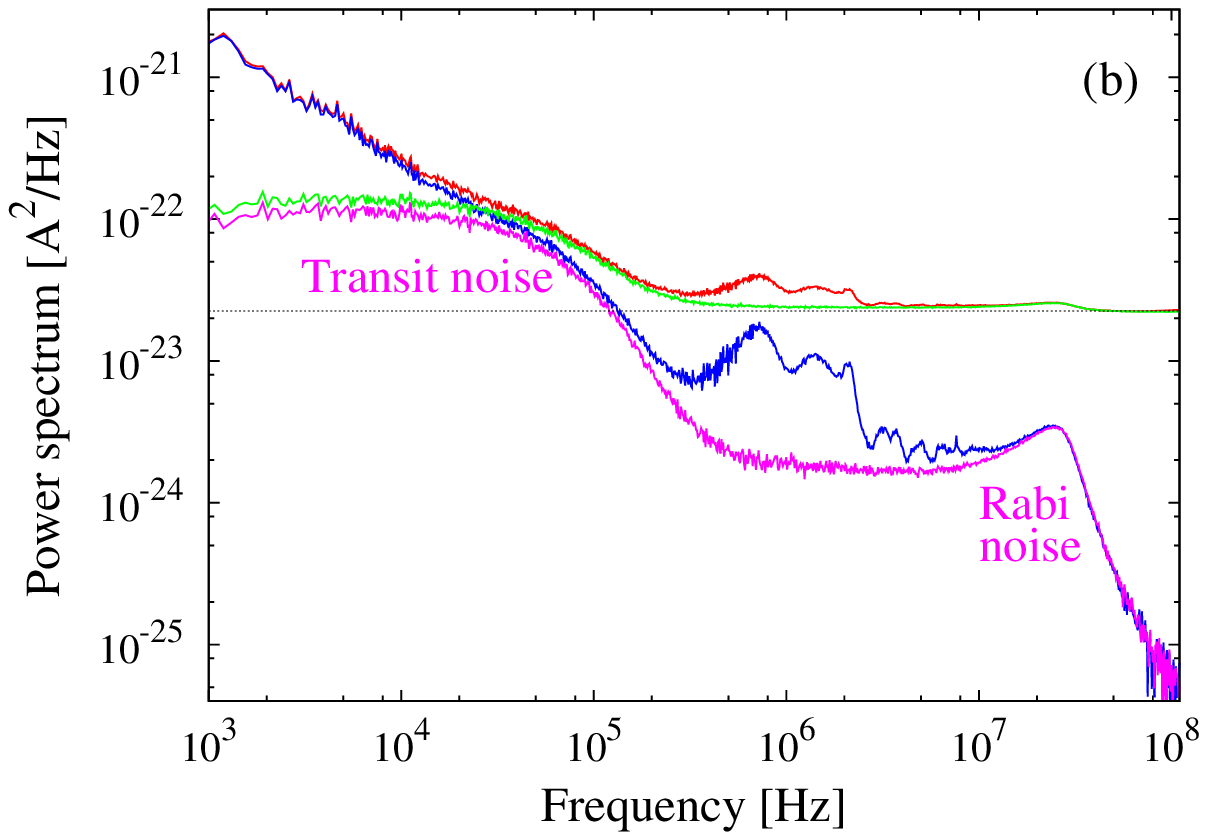}
    \caption{Spectra measured using the measurement systems \method
      a--\method d, for $P=28$ (a), $564\,\mu$W (b). 
      The notations are the same for both figures.
      \method a: Straightforward
      measurement with no noise reduction, \eqnn{spectrumA},
      \figno{setup}(a) (red). %
      \method b:      Measurement with differential detection, \eqnn{spectrumB},
      \figno{setup}(b) (green). %
      \method c: Measurement with shot noise reduction, \eqnn{spectrumC},
      \figno{setup}(c) (blue). %
      \method d: Measurement with both differential detection and shot noise
      reduction, \eqnn{spectrumD}, \figno{setup}(d) (magenta). %
      The shot noise level (black, dashed). }
    \label{fig:noiseReduction}
\end{figure}
We now examine, using experimental results, the performance of the
measurement system that includes noise reduction scheme and the
stabilized laser system explained above.  In the experiment, Rb atoms
sealed in vacuum Pyrex glass cells were used as samples and were
heated as necessary.  We transmitted circularly polarized light in
resonance with the ${}^{85}$Rb-D$_2$ transition from the hyperfine
level $5\rm {}^2S_{1/2}(F=3)$ to $5\rm {}^3P_{3/2}$\cite{RbSpecs}
through the cells.  Light source was circularly polarized using a
quarter-wave plate.

In \figno{noiseReduction}, measurements of the spontaneous noise
spectrum using the four measurement systems {\it a---d} in the
previous section, are displayed for beam waist $0.64$\,mm, cell temperature
47.4\,$^\circ$C and light powers $P=28,\ 546\,\mu$W.  
Let us briefly explain the physics underlying these
spectra\cite{amRb}.  The main components of this spontaneous noise
spectrum are the transit noise and Rabi noise.  Broad structure at
lower frequencies is the transit noise, caused by the effect of atoms
transiting the beam. The atoms travel at thermal velocities and the
average transit time for crossing the beam is $\sim 5\mu$s,
corresponding to a frequency, $f=2\times10^5$\,Hz, consistent with the
frequency this noise drops off. The higher frequency structure is
caused by the Rabi flopping of the atoms, in which the atom is excited
and de-excited by the incoming photons. Including the effects of the
spontaneous decay from the excited level, the gaussian nature of the
electric field strength within the light beam and Doppler shifts due
to atomic motion gives rise to the observed structure, which is not a
simple peak.
While not seen here, Zeeman noise due to Larmor precession of the
atoms can also be observed when a static magnetic field is
additionally applied.

In the observed spectra, it should be noted that the unwanted noise in
the laser and the detector system has been reduced to a level such
that differential detection has very little effect when the power is
weak (see \figno{noiseReduction}(a), $P=28\,\mu$W case, \method{a}
vs. \method{b}, \method{c} vs. \method{d}).
On the other hand, we can clearly see a reduction by a factor of
$10^3$ in the shot noise, achieved by using an averaged correlation of
the measurements (Eqs.~(\ref{eq:spectrumC}), (\ref{eq:spectrumD}),
\method{a} vs. \method{c}, \method{b} vs \method{d}), in both cases,
$P=28,\ 546\,\mu$W.
For the case $P=546\,\mu$W (\figno{noiseReduction}(b)), laser
noise induces substantial changes to the spectrum, which is clearly
eliminated using differential detection (\method{c}
vs. \method{d}). Optically coupling an external cavity to the
semiconductor laser destabilizes it weakly, resulting in the
contributions to the spectrum at frequencies around $1$\,MHz.  The
excess signal below 10\,kHz observed without differential detection
(\method{a}, \method{c}) also originates from the laser noise.  Looking
closely, even for the case $P=28\,\mu$W, small effects of differential
detection are visible below 10\,kHz and around 1\,MHz, due to the
reduction in the laser noise induced signals mentioned above.

The number of averagings is 
$ {\cal N}=T\Delta f$,
where $T$ is the {\it total} measurement time and $\Delta f$ is the
frequency resolution.  In the examples shown in
\figno{noiseReduction}, $\cal N$ is not uniform across the spectrum.
We used $\Delta f\geq 100\,$kHz for $f\gtrsim10\,$MHz and
$T\simeq10\,$s, so that ${\cal N}\gtrsim10^6$, leading to a noise
reduction factor of $1/\sqrt{\cal N}\lesssim10^{-3}$ for higher
frequencies, where the spectrum is the smallest. To obtain noise
reduction by a factor $10^{-4}$, $T\simeq1000\,$s is necessary for
$\Delta f=100\,$kHz. At lower frequencies, $\Delta f$ needs to be
smaller, so that $\cal N$ is also smaller and the shot noise is not
reduced as much. However, the spectrum has a larger value so that a
small noise reduction factor is also unnecessary.  At lowest
frequencies, the small number of samples is reflected in the slightly
jittery form of the observed spectrum.  Analog to digital conversion
was performed by PicoScope5203 (ADC, 8\,bit, sampling rate 500\,MHz,
PicoTechnology). Fourier transforms and averagings were computed on a
personal computer.  In practice, the actual time needed for the
experiment can be longer than $T$ due to two factors, the data
transfer rate of the ADC and the computation speed of the computer.

\section{Hyperfine level coherence effects}
\label{sec:hfCoherence}
\begin{figure}[htbp]
    \centering
    \includegraphics[width=\hnr cm,clip=true]{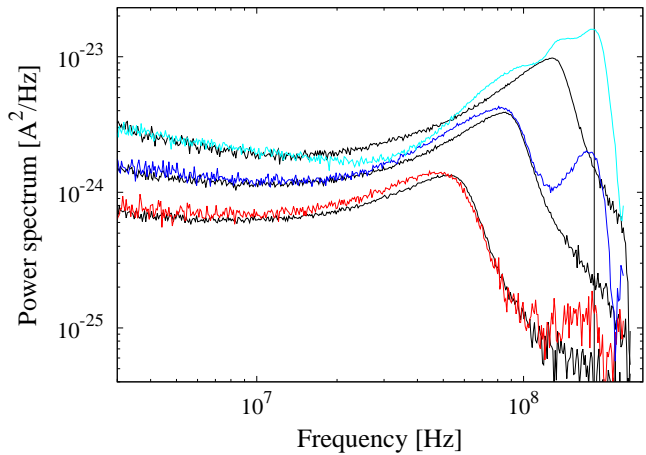}
    \caption{Spontaneous noise spectrum of ${}^{85}$Rb-D$_2$
      transitions when a light source with FM noise is used. $P=250 $
      (red), 552 (blue) and 1240\,$\mu$W (cyan). The vertical black
      line indicates $ F=2,4$ hyperfine splitting in the excited
      state, 184.0\,MHz. A clear peak structure at this frequency,
      irrespective of the light power, is seen for $P=552, 1240\,\mu$W
      cases, when the spontaneous noise is also appreciable at this
      frequency. The beam waist is $0.14$\,mm.
      The spontaneous noise spectrum obtained using a light source
      with approximately the same parameters, without the FM noise are
      also shown (black).  Their overall magnitude of the spectrum has
      been rescaled to make the comparisons clearer.}
    \label{fig:hfCoherence}
\end{figure}
When the spectral narrowing is insufficient, the applied laser light
contains FM noise. Such is the property of the stabilized laser output
2, which is not the output solely from the external cavity but
includes direct light from the laser diode.  The frequency modulation
in the light couples the hyperfine levels of the atoms in the
vapor. Since the hyperfine levels are essentially mixed, the states
evolve with the time dependence dictated by the hyperfine
splittings. This coherence of the hyperfine levels by itself is {\it
  not} visible in the fluctuation spectrum, since the frequency
modulation of the laser is common to all atoms and is canceled in the
differential measurement, system \method b in \secno{setup}. However,
this coherence affects Rabi noise and when it does so, it appears in
the spontaneous noise spectrum. In \figno{hfCoherence}, we can clearly
see the effects induced by the laser FM noise and its distinct
structure at a hyperfine level splitting frequency of the excited
state, regardless of the light power.  While not shown here, the
structure can be seen at the same frequency also when linearly
polarized light is applied. The effect is small when the spontaneous
noise is small, since this is necessary for the effect to appear in
the differential measurement, in addition to the coherence of the
hyperfine levels.  This interesting interplay of hyperfine level
mixing and spontaneous noise has not been seen previously, to our
knowledge.
\section{Conclusions and discussion}
\label{sec:conc}
We have developed a noise reduction system for measuring the
fluctuation spectra of transmitted light power, which allows us to
measure spectra at sub-shot noise levels. In the process, we also
devised a method for stabilizing the laser to achieve low levels of
both AM and FM noise.  With these methods, we measured the spontaneous
noise spectra down to levels $10^{-3}$ times the shot noise level in
this work and $10^{-4}$ times the level elsewhere\cite{amRb}.  
We further used the measurement system to observe a new phenomenon,
the spontaneous noise spectra for atomic level transitions with
coherence between hyperfine levels. 
The shot noise reduction is statistical, so that given enough measurement
time, the shot noise contribution can be reduced to arbitrarily low
levels. This reduction method relies on the uncorrelated nature of the
photon flux fluctuations in multiple measurements. The existence of
cross talk between detected signals, which can arise at low levels, is
one possible limitation in this regard.
Naturally, to achieve noise reduction, all extraneous noise needs to
be reduced to the desired level.  Another practical limitation in this
regard is the instability in the laser system.

Our measurement system with noise reduction is not much more difficult
nor expensive to set up than the simplest measurement scheme in
\figno{setup}(a) for obtaining the intensity fluctuation spectrum. We
expect such measurements of spontaneous noise to bring about a deeper
understanding of photon atom interactions, which underlie many ongoing
developments in physics, such as the physics of cold atoms, quantum
computing and atomic clocks. Our spectral measurements were performed
using atomic vapor cells, which are easy to handle and allows us also
to examine properties of buffered gas. By comparing the unbuffered
case to this, the physics with time scales longer than the buffered
gas collision time scale  is delineated, including transit noise
and Rabi noise\cite{amRb}. Our measurement system can also be applied
spectral measurements of atomic beams.

The spectral measurements of transmitted light power fluctuations
complement those of atomic fluorescence, where Rabi flopping was first
observed\cite{Mollow}. In comparison, transmitted light spectroscopy
has some advantages in that it has a higher power and is naturally a
heterodyne measurement so that the spectrum can be studied with a
larger dynamic range. Furthermore, the measurements can be more easily
set up, take less time and phase information can also be extracted
due to their heterodyne nature. 

\acknowledgments K.A. was supported in part by the Grant--in--Aid for
Scientific Research (\#20540279) from the Ministry of Education,
Culture, Sports, Science and Technology of Japan.

\end{document}